\documentclass[preprint,prb,aps,showpacs]{revtex4}   %Preprint version
\usepackage{epsfig}
\begin{document}
\def\fett #1{{\bf #1}}

\title{Magnetoresistance through spin polarized p-states}
\author{Nikos Papanikolaou}
\affiliation{Institute of Materials Science,
NCSR "Demokritos", GR-15310 Ag. Paraskevi Attiki, Greece}
\date{\today}

\begin{abstract}
We present a theoretical study of the ballistic magnetoresistance 
in Ni contacts using  
first-principles, atomistic electronic-structure calculations. In particular
we investigate the 
role of defects in the contact region in order to explain
the recently observed spectacular magnetoresistance ratio. 
Our results predict that possible presence of spin polarized  
oxygen in the contact region, could explain conductance changes 
by an order of magnitude. Electronic transport 
essentially occurs through spin-polarized oxygen p states, and this mechanism 
gives a much higher magnetoresistance than that obtained 
assuming clean atomically sharp domain walls alone.

\end{abstract}

\maketitle

%------------------------------------------------------------------
Understanding spin transport in order to use it in electronics devices 
is an intensively pursuited goal over the last 
years. Towards this 
direction spectacular effects were reported in electrodeposited 
Ni nanocontacts where resistance changes by two to three orders of magnitude  
when applying a small magnetic field at room 
temperature. \cite{ChHuPRB0203,WaChGa02} This has
followed after effort was focused on the magnetoresistance effect of
electrodeposited nanocontacts. \cite{Garcia0,Garcia} 
Smaller magnetoresistance
ratios, up to few hundred percent, were reported 
in ballistic Ni, Co, and Fe nanoconstrictions 
\cite{TaZhMuPRL99} and $Fe_3O_4$ nanocontacts. \cite{VeBaCoPRL01} 
There have been several attempts to explain the experimental data.
One possibility considered is the scattering of the electrons in a domain
wall. \cite{PhMuPRL02,TaVoEfPRB01,TaZhMuPRL99,VeBaCoPRL01} 
The width of a domain wall in a nanocontact
is predicted to be of the order of the nanocontact size, \cite{BrPRL99}
and such a reduction of the domain wall width increases the magnetoresistance.
Magnetostriction effects could also lead
to an increased magnetoresistance, \cite{VeBaCoPRL01} but most probably cannot
explain the experiments. \cite{Garcia0} 
Alternatively, a thin domain wall pinned in a magnetic dead layer
together with the possibility of 100\% polarization of the Ni d electrons was
also proposed as a possible explanation. \cite{WaChGa02,Garcia0} Moreover 
the effect of several ballistic nanocontacts in parallel requires further
study as pointed out in the most recent work of Ref. \onlinecite{ChHuPRB0203}.  

There is some experimental effort to measure the magnetoresistance due to a
domain wall in a constriction,
\cite{RuYuThPRB99,MiShMiJAP02,TaNaNaPRL99,EbRaHePRL00}
but all studies found magnetoresistance values of the order of 1\% or less.
This result is not far from theoretical predictions of Levy and Zhang, who
reported values between 2\% and 11\%. \cite{LeZhPRL97} 
Ab initio calculations for the ballistic conductance through an atomically
sharp domain wall \cite{HoScBrPRB99} predict MR ratios up to 60-70\%, 
which is at least one order of
magnitude lower than the experimental values on the electrodeposited Ni
nanocontacts. On the other hand, phenomenological 
models \cite{TaVoEfPRB01,VeBaCoPRL01} seem to account for   
magnetoresistance values up to a few hundred percent, 
by assuming domain wall scattering. However it is difficult to assign a
realistic spin polarization in order to explain 
magnetoconductance (MC) values of the order of
thousands of per cent.   

Since the big effects appear on electrodeposited Ni, nanocontacts 
\cite{Garcia}
the electrodeposition process should play a decisive
role for the nanocontact properties. The chemistry and structure of the
contacts is not yet clear, while there is some indication that the diameter of
the nanocontacts is not the crucial factor to obtain big magnetoresistance.
\cite{WaChGa02}
Generally, different conductance values are attributed to different diameters 
of the contact area, however the transport mechanism is still not clear. 

In this work
we propose a new mechanism, based on the presence of 
spin-polarized oxygen in
the contact region, that could explain rather high magnetoresistance  values 
in Ni contacts. We considered Cl, S, O, and C in the contact 
region \cite{WaChGa02}, 
and studied the influence of the nanocontact chemistry on
the magnetotransport properties by means of electronic structure
calculations. Only oxygen was found to have a magnetic moment in the
structures we studied, while a presence of a magnetic dead layer cannot
explain the observed effect since it reduces the magnetoresistance ratio and
this is confirmed by our calculations. 

We model the nanocontacts by two semiinfinite Ni (001) ideal   
leads covered with a full monolayer (ML) of Cl, S, C and O, 
and bring them close together. We use the experimental lattice constant of Ni: 
$a_0$=6.65 au. The distance of the defect layer 
from the Ni surface is $a_0/2$
and the defect layers are separated by $a_0$.
We do not intent to model an O overlayer on the Ni (001) surface, the
structure and magnetic properties of O on Ni is studied well in the past
both theoretically and experimentally, \cite{OonNi,StWaAlJCP97} 
and is not relevant to the
structure of electrodeposited Ni nanocontacts. Our aim is to use a 
simple model structure in order to achieve the
minimum requirements that would demonstrate the effect of spin polarized
transport through p-states. The experimental structure in the nanocontacts
where the high magnetoresistance is observed is not well known, but
spin polarization on the O atoms can occur for example if a thin NiO layer is
present in the contact. \cite{NiOO} Moreover we assume that the electronic
transport is ballistic. 

The electronic structure is calculated using the Screened
Korringa-Kohn-Rostoker (KKR) Green's function method, 
details of which can be found elsewhere.
\cite{PaZeDeJPB02} The potential is assumed to be spherically symmetric
around each atom. We considered an angular 
momentum cutoff $l_{max}=3$ for the
wavefunctions and the Green's function, and $l_{max}=6$ in the multipole
expansion  of the charge density. The screening transformation
of the KKR method allows for the description of semi-infinite 
geometries using the decimation method. 
Exchange and correlation
effects are taken into account within the local density 
approximation of the density
functional theory \cite{LDA} using the parameterization of Vosko, 
Wilk, and Nusair. \cite{VoWiNuCJP80} 
The conductance $g$ is calculated within the Kubo linear response theory 
as formulated by Baranger
and Stone. \cite{BaStPRB89} We consider a sample connected to 
two semi-infinite leads. Two
surfaces $C_n$ and $C_m$ separate the leads from the sample. Following
Ref.~\onlinecite{BaStPRB89} the conductance between the two leads at energy 
$E$ and at zero temperature is given by
\begin{equation}
g = \frac{-e^2 {\hbar}^3}{8\pi M^2}
\int_{C_n}{dS \int_{C_m}{dS' G^{+}({\mathbf r},{\mathbf r'};E)
\stackrel{\leftrightarrow}{\nabla}
\stackrel{\leftrightarrow}{\nabla}' 
G^{-}({\mathbf r'},{\mathbf r};E)}}
\label{cond1}
\end{equation} 
where $ f(r) \stackrel{\leftrightarrow}{\nabla}g(r) = 
      f(r)\nabla g(r) - 
      g(r)\nabla f(r)$, and $G^+$, $G^-$ are the advanced and
retarded Green's functions, respectively. 
The systems under consideration have a two-dimensional (2D) periodicity
parallel to the interfaces, and the surface integrals in Eq.~(\ref{cond1}) 
are performed using a Fourier transform. 
A detailed presentation will be
reported elsewhere. \cite{MaPaDe02} 
The Green's function for the conductance
calculation was evaluated using a small imaginary part to 
the energy ImE=0.02 mRy.
  
\begin{figure}
  \begin{center}
    \epsfig{file=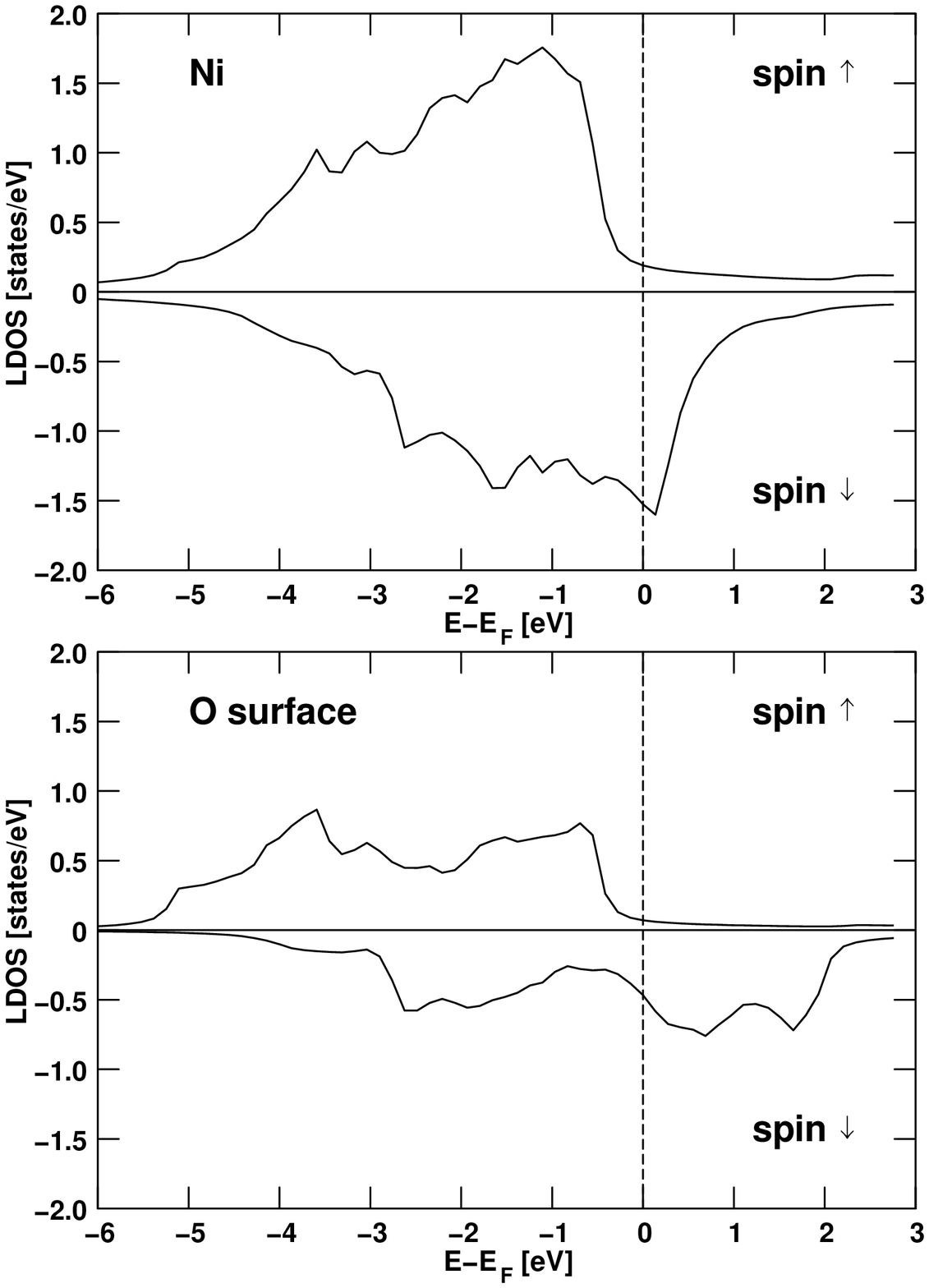,width=2.9in,angle=0}
    \caption{Spin resolved density of states for the interface  
O and first Ni layers for the 1 ML oxygen on fcc Ni (001) 
junction.}
    \label{fig1}
  \end{center}
\end{figure}

\begin{figure}[tb]
  \begin{center}
    \epsfig{file=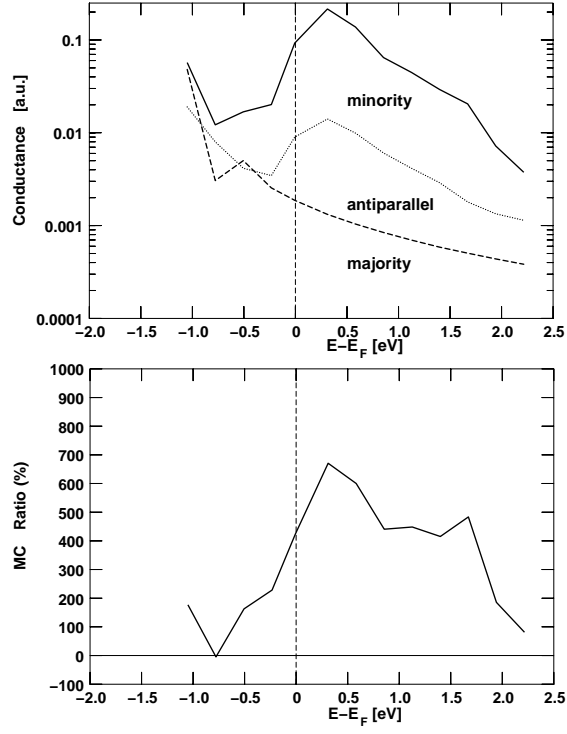,width=2.9in,angle=0}
    \caption{Top panel: Conductance variation with energy for the 1ML oxygen on
Ni (001) junction, for majority (full line) and minority (dashed line) 
spin channels in the parallel spin configuration together 
with the antiparallel spin configuration (dotted line). Conductance in
arbitrary units. Bottom
panel: the magnetoconductance ratio for the same junction.}
    \label{fig2}
  \end{center}
\end{figure}

In Fig. \ref{fig1} we present the local density of states (LDOS) 
of O at the interface and of Ni at the neighboring layer. The
magnetic moment of Ni is 0.85 $\mu_B$ compared to the bulk value
of 0.61 $\mu_B$; this value is very close to the clean Ni (001) surface
moment. The oxygen in our structure has a rather big moment of
1.4 $\mu_B$. We note that the difference of 
the LDOS at the Fermi level is larger for the Ni (d) states compared with the 
(p) states of O as can seen in Fig. \ref{fig1}. 
However, the induced p magnetization is the key factor to
achieve high magnetoconductance ratios. In Fig.~\ref{fig2} we present the
conductance for each spin channel, for parallel (P) and antiparallel (AP)
orientation of the magnetic moments in the Ni leads, together with the 
MC ratio, defined as $\mathrm{MC} = \frac{g_P-g_{AP}}{g_{AP}}$. 
The magnetic profile
changes only little in the antiparallel configuration and O remains highly
spin polarized. As we can see in the bottom panel of Fig. \ref{fig2} the MC
ratio was found to be about 450\% at the Fermi level and increases up to almost
700\% at 0.3 eV above $E_F$. The conductance and MC ratio peaks are closely
related to the LDOS at the O site. Comparing Fig. \ref{fig1} with
Fig. \ref{fig2}, we can see that since the majority-spin states are occupied 
(they are below $E_F$), the variation of the MC
ratio above $E_F$ reflects the variation of the minority LDOS. The p
polarization leads to a large contrast in the conductance between majority and
minority channels. Although the spin polarization of Ni at $E_F$ is
almost twice larger than the O one, it is the relatively small spin
polarization of the oxygen p states which is responsible for the spectacular
MC ratio.
Indeed, our calculations show that in a pure Ni
contact where two Ni (001) leads are separated by one vacuum layer the MC
ratio has a maximum value of only $\approx 70 \%$, which is 
similar to the values reported previously for the magnetoresistance 
of abrupt, atomically sharp domain walls in the 
ballistic regime \cite{HoScBrPRB99}. 
The question of spin polarization and its relevance to the observed 
current spin polarization has been discussed before. \cite{Mazin01} Our
results introduce another factor, the character of the
polarization, which is missing from the models used previously 
to explain the observed effects. \cite{TaVoEfPRB01,VeBaCoPRL01}
Spin polarization of p orbitals is 
more difficult to occur but, once
this is achieved, highly spin-polarized currents are possible.
In the present work this was demonstrated with O but other    
substances might lead to similar effects as well.

In order to exclude the possibility that the big MC values are an artifact of
the 2D periodicity we used a c(2x2) structure for the O on the surface of Ni
(001) and recalculated the transport properties. Despite the fact that 
c(2x2) structure is close to the experimental structure for an O overlayer on
Ni (001), \cite{StWaAlJCP97} the distance 
between Ni and O we used is almost
double compared with the experimental structure 
in order to achieve polarized O. Since transport is mainly 
determined by the contact region we are interested just on 
the effect of spin-polarized O on a magnetic substrate.  

\begin{figure}[tb]
  \begin{center}
    \epsfig{file=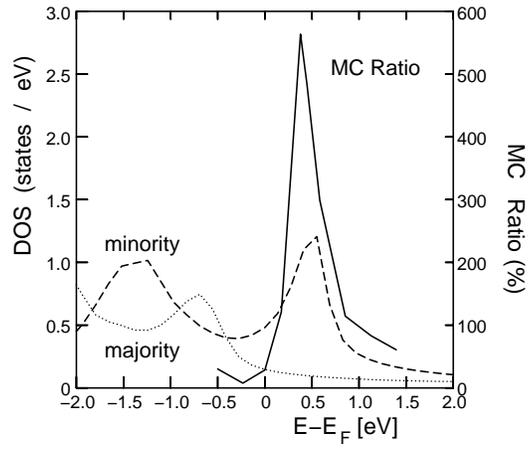,width=2.9in,angle=0}
    \caption{Spin resolved density of states on the O site for 
two Ni (001) leads covered with 0.5 ML of oxygen on a c(2x2) structure 
and separated with one vacuum layer. The
magnetoconductance ratio as a function of energy close 
to the Fermi level is also shown (full line).}
     \label{fig3}
\end{center}
\end{figure}

Our results for the LDOS close to the Fermi
level together with the MC ratio are shown in Fig. \ref{fig3}. The magnetic
moment on the O atoms is now reduced to 0.95 $\mu_B$, while the Ni moment is
less affected. This is also seen in the LDOS of O where 
the minority-spin p states 
are lower in energy compared to the case of one ML oxygen coverage.
As we can see
the large MC effect survives also in this configuration. 
The MC ratio is rather small
at $E_F$ but increases rapidly just 0.5 eV above $E_F$. The largest MC 
values, of about 500\%, are obtained at the oxygen p peak position. 
In other words, p orbitals can
build highly conducting channels and, if they are spin polarized, conductance
in one spin channel can be even two orders of 
magnitude higher than that in the 
other spin channel. However in the systems we studied the MC ratios 
are limited to $~700\%$ since conductance is not so drastically reduced in the
antiparallel configuration. 
We note that the proposed mechanism: high current when 
a bond is formed, vanishingly small current when the bond is 
broken, in the antiparallel configuration, cannot exclude even 
higher MC ratios depending on the exact structure chosen. 
However it is important that the O-O bond energy should be smaller than 
exchange energy so that the bond can
break and reform by applying a magnetic field. 
The interaction between the
two magnetic leads should be weak enough so that the domain wall 
is pined at the contact region.
 
Despite the fact that the proposed mechanism can explain high
magnetoconductance values it cannot fully account for the experimental
findings. In particular experimentally both positive and negative
magnetioresistance is observed, \cite{ChHuPRB0203} moreover conductance 
is generally big at
zero magnetic field and becomes much smaller when applying a magnetic field.
The influence of an antiferromagnetic NiO layer at the contact region requires
further study, but as the present work shows large changes in 
conductance can result from spin polarized O-O bonds.

It is important that a big magnetic moment on the O atoms is not a
necessary  condition to obtain big MC ratio. Only a big difference
in the LDOS between minority- and majority-spin electrons at the Fermi 
level is required.  
Moreover Ni itself is not a necessary ingredient; any
magnetic material that could support O at the contact region and induce spin
polarization to the O atoms would have similar properties. 
 
Finally we would like to make a clear distinction 
between the phenomena presented here
and the effect of hot spots in the conductance studied previously in
ferromagnet/vacuum/ferromagnet tunnel junctions. 
\cite{WuPaZe02} In that case tunneling
coexists with transport through interface states, which have a transmission
coefficient close to unity and are localized in a very 
small region of the surface 
Brillouin zone. However the hot spots depend sensitively on 
the symmetry of the junction
and require flat, ideal interfaces. Any small disorder would suppress the
effect. On the contrary, the big MC ratios reported here are related to 
spin-dependent transport through O-O bonds. The 2D structure is not a crucial
factor and this is also clear from the $k_{\parallel}$-dependent conductance
where the variation is relatively smooth and a big part of the 
surface Brillouin
zone gives significant contributions. 

Summarizing,  
our results show that high MC ratios in Ni nanocontacts, 
could be explained by spin-dependent defect scattering at the
nanocontact. This is in agreement with the conclusion of Ref.
\onlinecite{LeZhPRL97} that defect scattering can enhance 
the magnetoresistance 
in a domain wall; however the role of spin-polarized p orbitals 
has not been pointed out up to now, to
our knowledge. We have demonstrated that if oxygen at the contact region is 
spin polarized, this gives rise to high MC ratios. This idea is not 
in contradiction with the idea of scattering in an atomically sharp domain
wall. However ab-initio ballistic transport calculations cannot explain
magnetoresistances of few hundred percent, assuming a defect-free atomically
sharp domain wall. On the contrary, the transport mechanism proposed in this
work can justify spin polarization
close to 100\% that is required to explain the spectacular MC ratios in
electrodeposited nanocontacts.
We hope that our theoretical
prediction will stimulate further experimental studies to clarify the role of 
chemistry and in particular of NiO on the magnetotransport
properties of nanocontacts.

\begin{acknowledgments}
I would like to thank O. Cespedeo for useful correspondence.
\end{acknowledgments}
%------------------------------------------------------------------

\newpage

\end{document}